\documentclass{article}
\usepackage{lnfprep}
\usepackage{graphicx}
\usepackage{subfigure}
\usepackage{amsmath}
\usepackage{amssymb}

\def\beq{\begin{equation}}   \def\eeq{\end{equation}}
\def\bea{\begin{eqnarray}}   \def\eea{\end{eqnarray}}

\newcommand{\gsim}{\lower.7ex\hbox{$ \;\stackrel{\textstyle>}{\sim}\;$}}
\newcommand{\lsim}{\lower.7ex\hbox{$ \;\stackrel{\textstyle<}{\sim}\;$}}

\def\c2{CLEO~II.V}

\def\d0d0{ D^0\bar{D}^0 }
\def\p0p0{ P^0\bar{P}^0 }

\def\qp2{ \Bigl| \frac{q}{p} \Bigr|^2 }
\def\pq2{ \Bigl| \frac{p}{q} \Bigr|^2 }

\def\ps2s{  \psi(2S) }
\def\q2{ $q^2$ }
\def\cm2s1{ $\,{\rm cm}^{-2} {\rm s}^{-1}$}
\def\d0{D_2^{*0}}
\def\d+{D_2^{*+}}

\newcommand{\Header}{
  \begin{tabular}{rl}
  \hspace{-.4cm}
      &
    \renewcommand{\arraystretch}{0.5}
    \renewcommand{\arraystretch}{1}
  \end{tabular}
  \vskip 1cm
  \begin{flushright}
  \renewcommand{\arraystretch}{0.5}
    \begin{tabular}{r}
      {\underline{INFN-14-22/LNF}}\\    
      {\small December 25th, 2014} \\      
    \end{tabular}
  \end{flushright}
  \renewcommand{\arraystretch}{1}
  \vskip 1 cm
  }

\begin{document}
\begin{titlepage}
\title{
  \Header{
    \LARGE 
     \textsc{
        \textmd{
  Performance of the gas gain monitoring system of the CMS
	RPC muon detector
   }
  }
 }
\author{
	L.~Benussi$^a$\thanks{Corresponding author.},
	S.~Bianco$^a$,
	L.~Passamonti$^a$,
	D.~Piccolo$^a$,
	D.~Pierluigi$^a$,
	G.~Raffone$^a$,
	A.~Russo$^a$,
	\\
	G.~Saviano$^{a,+}$,
	Y.~Ban$^b$,
	J.~Cai$^b$,
	Q.~Li$^b$,
	S.~L$^b$iu,
	S.~Qian$^b$,
	D.~Wang$^b$,
	\\
	Z.~Xu$^b$,
	F.~Zhang$^b$,
	Y.~Choi$^c$,
	D.~Kim$^c$,
	S.~Choi$^d$,
	B.~Hong$^d$,
	J.W.~Kang$^d$ , 
	\\
	M.~Kang$^d$,
	J.H.~Kwon$^d$,
	K.S.~Lee,$^d$,
	S.K.~Park$^d$,
	L.~Pant$^e$,
	V.B.J.~Singh$^f$,
	A.M.R.~Kumar$^f$,
	\\
	S.~Kumar$^f$,
	S.~Chand$^f$,
	A.~Singh$^f$,
	V.K.~Bhandari$^f$,
	A.~Cimmino$^g$,
	A.~Ocampo$^g$,
	F.~Thyssen$^g$,
	\\
	M.~Tytgat$^g$,
	W.~Van Doninck$^h$,
	A.~Ahmad$^i$,
	S.~Muhamma$^i$,
	M.~Shoaib$^i$,
	H.~Hoorani$^i$,
	I.~Awan,$^i$
	\\
	I.~Ali$^i$,
	W.~Ahmed$^i$,
	M.I.~Asqhar$^i$,
	H.~Shahzad$^i$,
	A.~Sayed$^l$,
	A.~Ibrahim$^l$,
	S.~Ali$^l$,
	\\
	R.~Ali$^l$,
	A.~Radi$^l$,
	T.~Elkafrawi$^l$,
	A.~Sharma$^m$,
	S.~Colafranceschi$^m$,
	M.~Abbrescia$^n$,
	\\
	P.~Verwilligen$^n$,
	S.~Meola$^o$,
	N.~Cavallo$^o$,
	A.~Braghieri$^p$,
	P.~Montagna$^p$,
	C.~Riccardi$^p$,
	\\
	P.~Salvini$^p$,
	P.~Vitulo$^p$,
	A.~Dimitrov$^q$,
	R.~Hadjiiska$^q$,
	L.~Litov$^q$,
	B.~Pavlov$^q$,
	P.~Petkov$^q$,
	\\
	A.~Aleksandrov$^r$, 
	V.~Genchev$^r$,
	P.~Iaydjiev$^r$,
	M.~Rodozov$^r$,
	G.~Sultanov$^r$,
	M.~Vutova$^r$,
	S.~Stoykova$^r$,
	\\
	H.S.~Ibarg\"uen$^s$,
	M.I.~Pedraza Morales$^s$, 
	S.~Carpinteyro Bernardino$^s$,
	I.~Bagaturia$^t$,
	Z.~Tsamalaidze$^t$.
} 
\\
} 
\maketitle
\baselineskip=1pt

\begin{abstract}
\indent 
  The RPC muon detector of the CMS experiment at the LHC (CERN, Geneva, Switzerland)
	is equipped with a Gas Gain Monitoring (GGM) system. A report on the
	stability of the system during the 2011-2012 data taking run is given, as
	well as the observation of an effect which suggests a novel method for
	the monitoring of gas mixture composition.
\end{abstract}

\vspace*{\stretch{2}}

\vskip 1cm
\begin{flushleft}
Submitted to JINST.  Keywords: CMS; RPC; muon detector; gas detector\\
\vskip 1cm
\begin{tabular}{l l}
  \hline\\
  \llap{$^a$}Laboratori Nazionali di Frascati dell'INFN,
	 Frascati, Italy \\
	\llap{$^b$}Peking University, Peking, China  \\
	\llap{$^c$}SKK University, Seoul, S.~Korea  \\
	\llap{$^d$}Korea University, Seoul, S.~Korea  \\
	\llap{$^e$}BARC, Mumbai, India\\
	\llap{$^f$}Chandigarh University, India \\
	\llap{$^g$}Ghent University, Ghent, Belgium \\
	\llap{$^h$}VUB, Bruxelles, Belgium \\
	\llap{$^i$}NCP, Islamabad, Pakistan \\
	\llap{$^l$}ENHEP, Cairo, Egypt  \\ 
	\llap{$^m$}CERN, Geneve,  Switzerland\\
	\llap{$^n$}Universit\`a di Bari and INFN Bari, Bari, Italy \\ 
	\llap{$^o$}Universit\`a di Napoli and INFN Napoli, Napoli, Italy  \\ 
	\llap{$^p$}Universit\`a di Pavia and INFN Pavia, Pavia, Italy \\ 
	\llap{$^q$}University of Sofia, Sofia, Bulgaria\\
	\llap{$^r$}INRNE, Sofia, Bulgaria  \\ 
	\llap{$^s$}Puebla University, Puebla, Mexico \\ 
	\llap{$^t$}Tbilisi University and IHEPI, Tbilisi, Georgia\\ 
	$~^+$also at Universit\`a degli Studi di Roma La Sapienza, Rome,
	Italy\\
	$~^*$E-mail: {\tt luigi.benussi@lnf.infn.it}
 \\
\end{tabular}
\end{flushleft}
\end{titlepage}
\pagestyle{plain}
\setcounter{page}2
\baselineskip=17pt

	\section{Introduction}\label{sec:INTRO}
	
	The Resistive Plate Counter (RPC) muon detector of the Compact Muon Solenoid (CMS)
	 experiment at the  Large Hadron Collider (LHC) at CERN
	is equipped with a Gas Gain Monitoring (GGM) system. Detailed descriptions of the GGM
	can be found in  
	\cite{Colafranceschi:2012qf, Colafranceschi:2010zz, Benussi:2008vs, Benussi:2008fp}.
	The system has been in operation for the whole duration of the
	2011-2012 CMS data taking period.
	A report of both its
	performance and the experience gained with it is given. 
	An effect which may suggest a novel method for the monitoring of gas mixture
	composition is discussed.
	\section{Experimental setup}\label{sec:SETUP}
\begin{figure}[Ht] 
	\centering
	\includegraphics[width=0.7\textwidth]{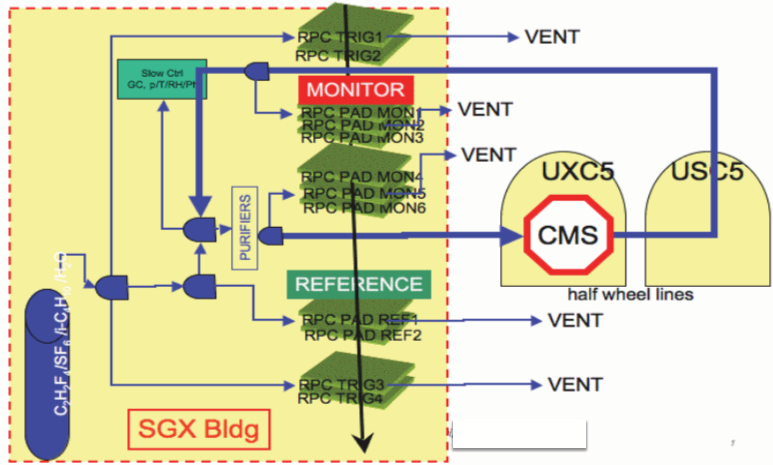}
	\caption{Schematic layout of the GGM and closed-loop gas recirculation
	system of the CMS RPC detector.}
	\label{fig:SETUP}
	\end{figure}

	The GGM system is
	composed of 12
	square single-gap RPC detectors arranged as a cosmic ray hodoscope 
	(Fig.\ref{fig:SETUP}). The GGM is located on the surface, in the SGX5 gas
	 room of the CMS
	experimental area.
	\par
	The system 
	is designed to provide a fast and accurate determination of any shift in the
	working point of its chambers due to gas mixture changes. 
	It compares three different gas mixtures from the CMS experiment: 
	a newly supplied fresh gas
	mixture, and the gas mixture
	before and after the filters of the closed-loop recirculation system. 
	The closed loop is fed with one volume ($14\,{\rm m}^3$) gas exchange per hour
	\cite{Capeans:2013uma}.  
	\par
	
	\begin{figure}[Hb] 
	\centering
	\includegraphics[width=0.9\textwidth]{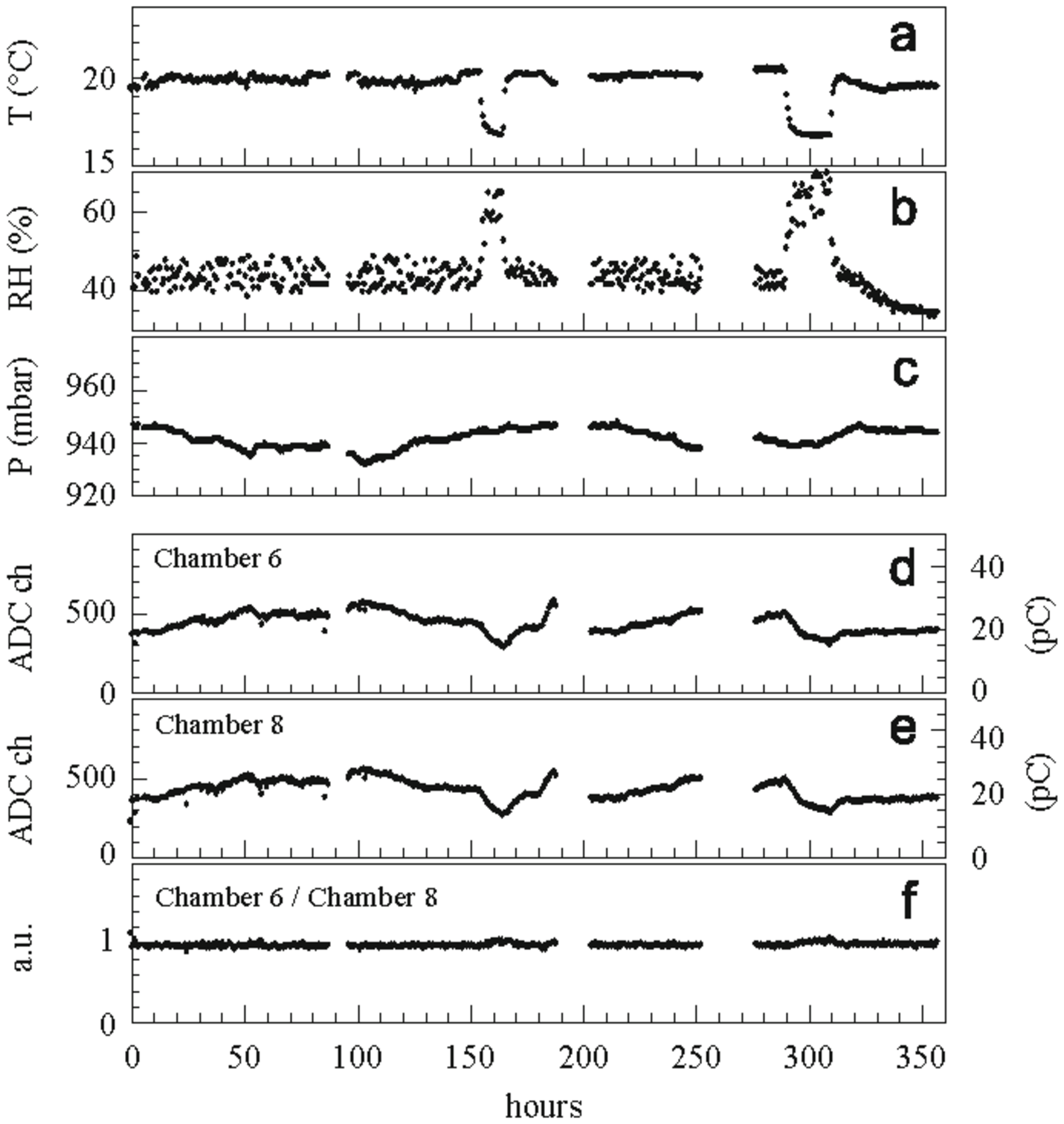}
	\caption{The principle of automatic compensation of environmental effects by
	charge ratios: 
	environmental variables such as temperature (a), relative humidity (b), atmospheric
	pressure (c), gaps anodic average charges (d,e) and their ratio (f).
	}
	\label{fig:GGMCANCELOUT}
	\end{figure}
	\par
	 The 12 single bakelite  gaps have a double pad
	 readout (4 trigger gaps, 8 signal gaps) which allows 
	 removal of coherent noise ({\it i.e.,} the environmental baseline
	 noise showing on both pads) by subtracting
	 algebraically signals from both pads.
	 The average event rate is 5 Hz, corresponding to about 30 minutes for 10000
	 events. Anode charge distributions are collected for blocks 
	of 30 minutes,
	 and changes of the charge averages over time provide indications of changes
	 in the working point of the GGM chambers. An automatic compensation of 
	environmental effects on the chamber responses is achieved by means of 
	two-gap ratios of anode charge distributions (Fig.\ref{fig:GGMCANCELOUT}) 
	\cite{Colafranceschi:2010zz}.
	\subsection{High-voltage scans}\label{sec:HVSCAN}
	The GGM system was designed to detect changes in the working point of the
	chambers via
	differences between a fresh gas mixture, and the mixtures before and after the
	purifiers of the gas recirculation system, using by also the charge ratio
	algorithm. 
	Next to that, weekly RPC high voltage (HV) 
	scans were performed which additionally
	provided a direct measurement of the GGM working point, allowing to spot
	changes in the gas mixture composition.
	
\begin{figure}[Hb] 
	\centering
	\includegraphics[width=0.8\textwidth]{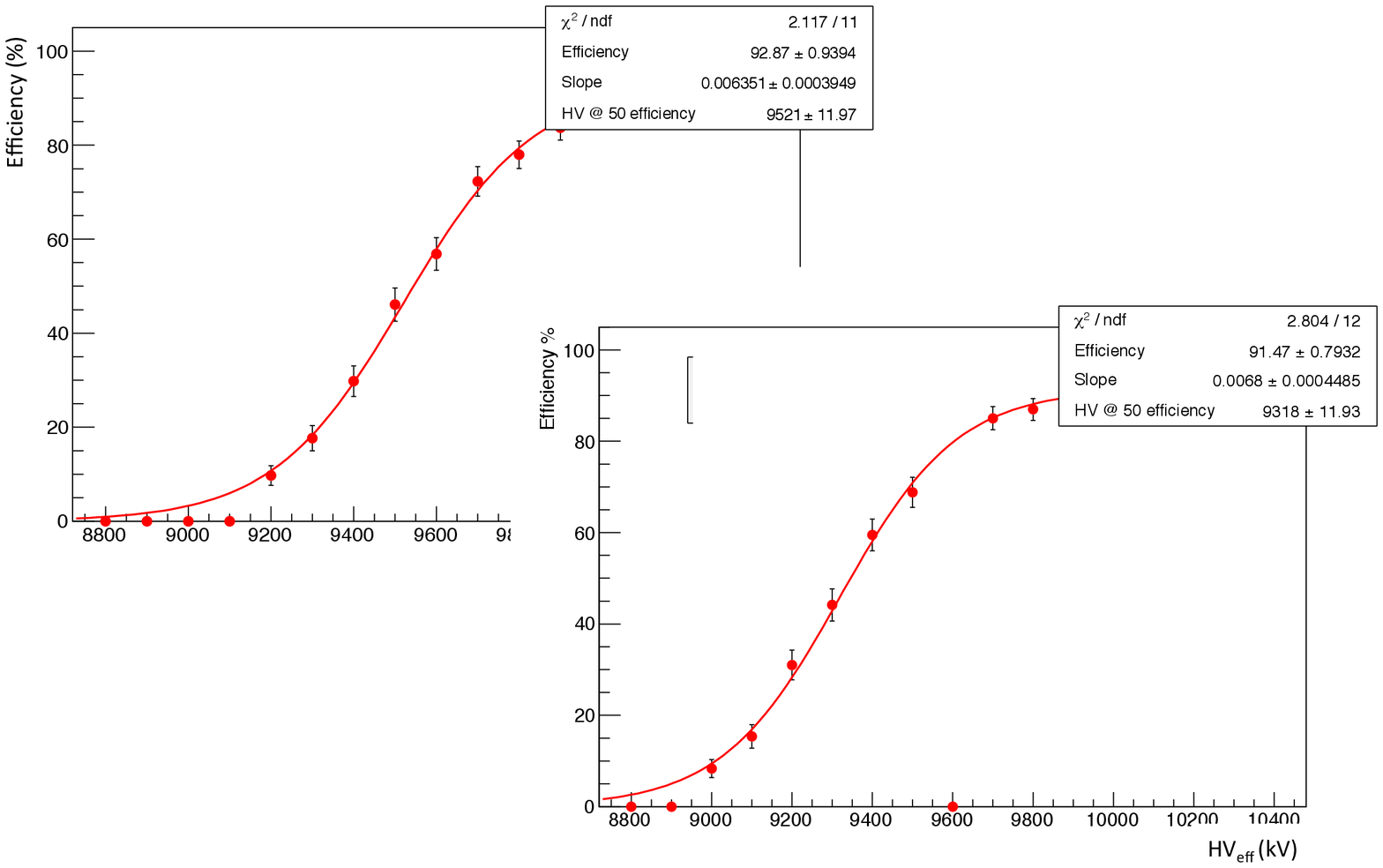}
	\caption{Typical HV scans for two GGM single-gap chambers.}
	\label{fig:HVSCAN}
	\end{figure}

	Fig.~\ref{fig:HVSCAN} shows typical HV scans of two GGM single-gap chambers,
	 as a function of effective HV supply 
	 \begin{equation} \label{eq:hveff}
	    HV_{eff} = HV\frac{p_0}{p} \frac{T}{T_0}.
	 \end{equation}
	 
	The $HV_{eff}$ value corresponding to the effective voltage where the chamber efficiency
	$\epsilon$
	is at  
	50\% of its maximum 
	\begin{equation} \label{eq:hv50}
	HV50 \equiv HV_{eff} (\epsilon = \epsilon_{max}/2)
	\end{equation}
	is the most sensitive parameter  to any change of the working point. 
	
\begin{figure}[Ht] 
	\centering
	\includegraphics[width=0.9\textwidth]{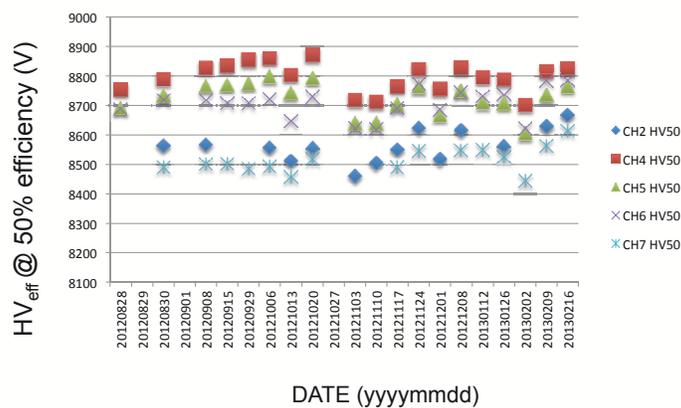}
	\caption{Dependence of HV50 (defined in text) 
	 as a function of time over the 2011-2012 data taking period.}
	\label{fig:HV50TIME}
	\end{figure}

	The HV50 parameter is shown for a few GGM chambers in Fig.~\ref{fig:HV50TIME} 
	as a function of time for the 2011-2012 data taking period.

	\begin{figure}[Hb] 
	\centering
	\includegraphics[width=0.8\textwidth]{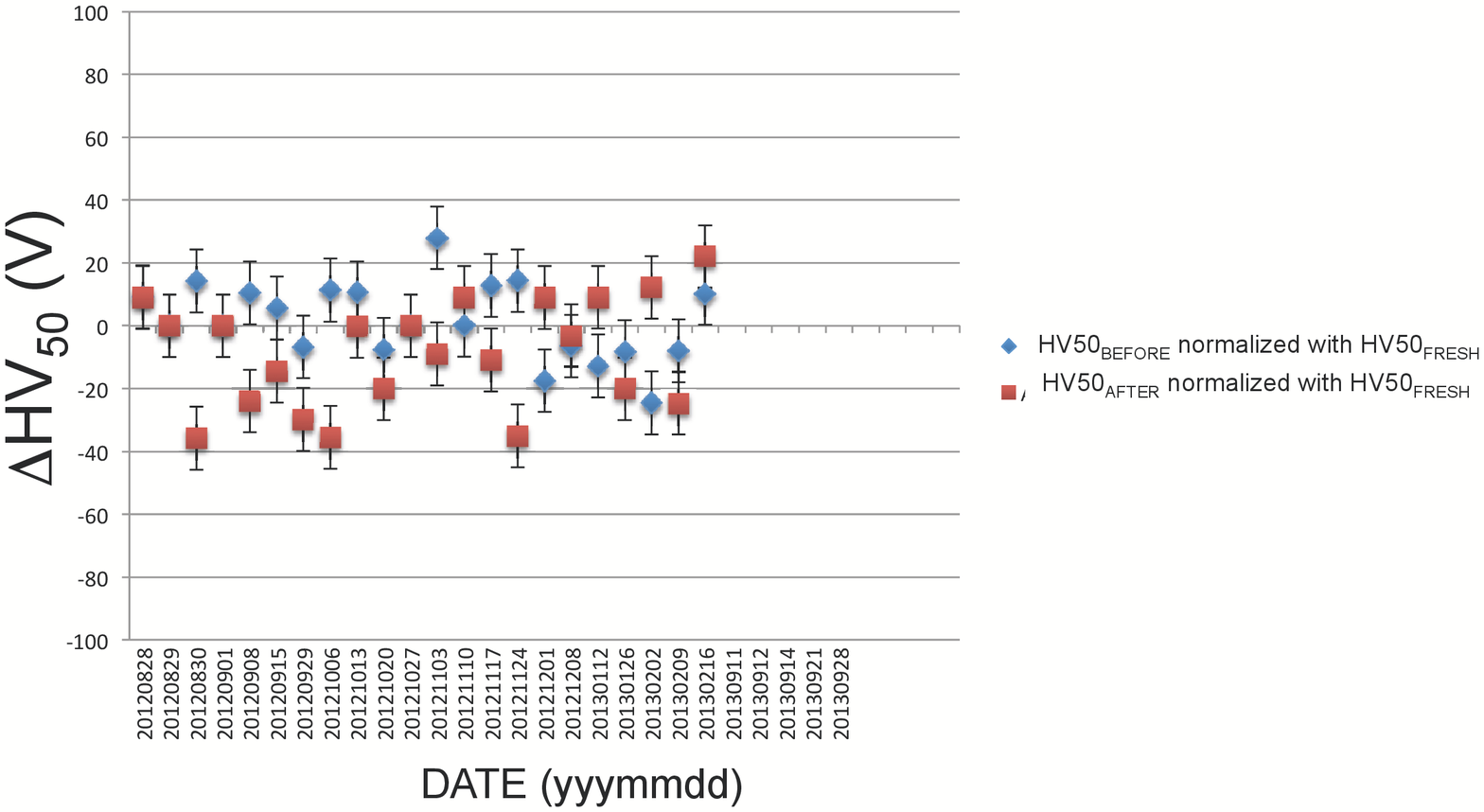}
	\caption{Dependence of normalized HV50.}
	\label{fig:DELTAHV50TIME}
	\end{figure}

	Each HV50 value is sensitive to common environmental 
	effects that are not corrected for by the HV feedback. 
	A normalization method taking the ratio of the HV50 values for the gas
	mixtures in the closed-loop system and the fresh gas mixture, provides
	a stable response over a period of several months 
	(Fig.\ref{fig:DELTAHV50TIME}). 
	$\Delta HV50$ is the difference in the HV50 parameter for the before- or after-type
	gaps normalized to the fresh-type gaps. The distribution of $\Delta HV50$ parameter
	over time is compatible with statistical errors and does not suggest any systematic
	effect that may indicate a possible problem on the gas mixture composition. 
	\subsection{Time transient}\label{sec:TTRANS}
\begin{figure}[Htbp] 
	\centering
	\includegraphics[width=.9\textwidth]{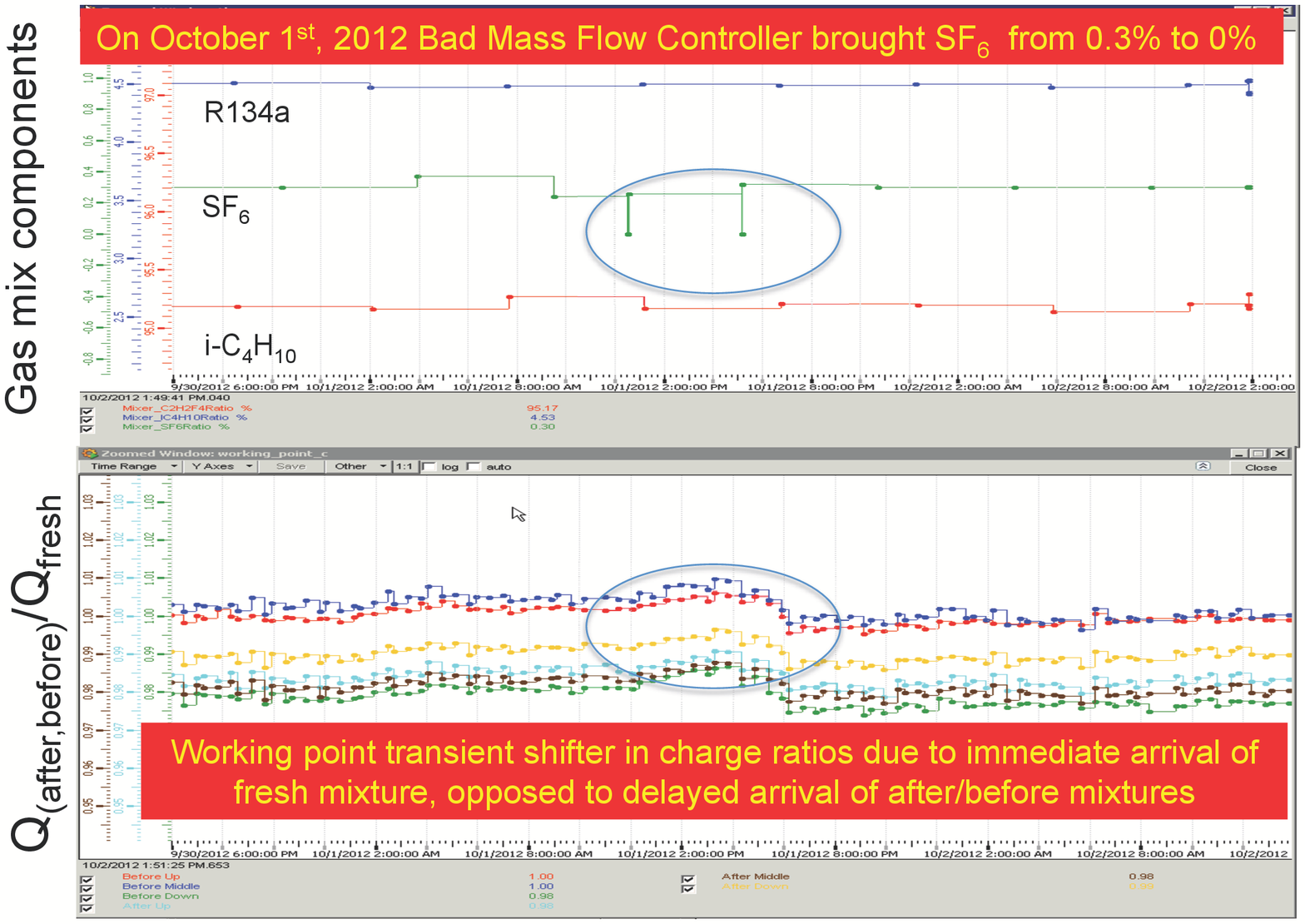}
	\caption{Failure of $SF_6$ mass flow controller (top); interference pattern observed in
	the online monitoring tool for
	all charge ratios of the GGM.}
	\label{fig:TRANSIENTONLINE}
	\end{figure}

\begin{figure}[Ht] 
	\centering
	\includegraphics[width=.9\textwidth]{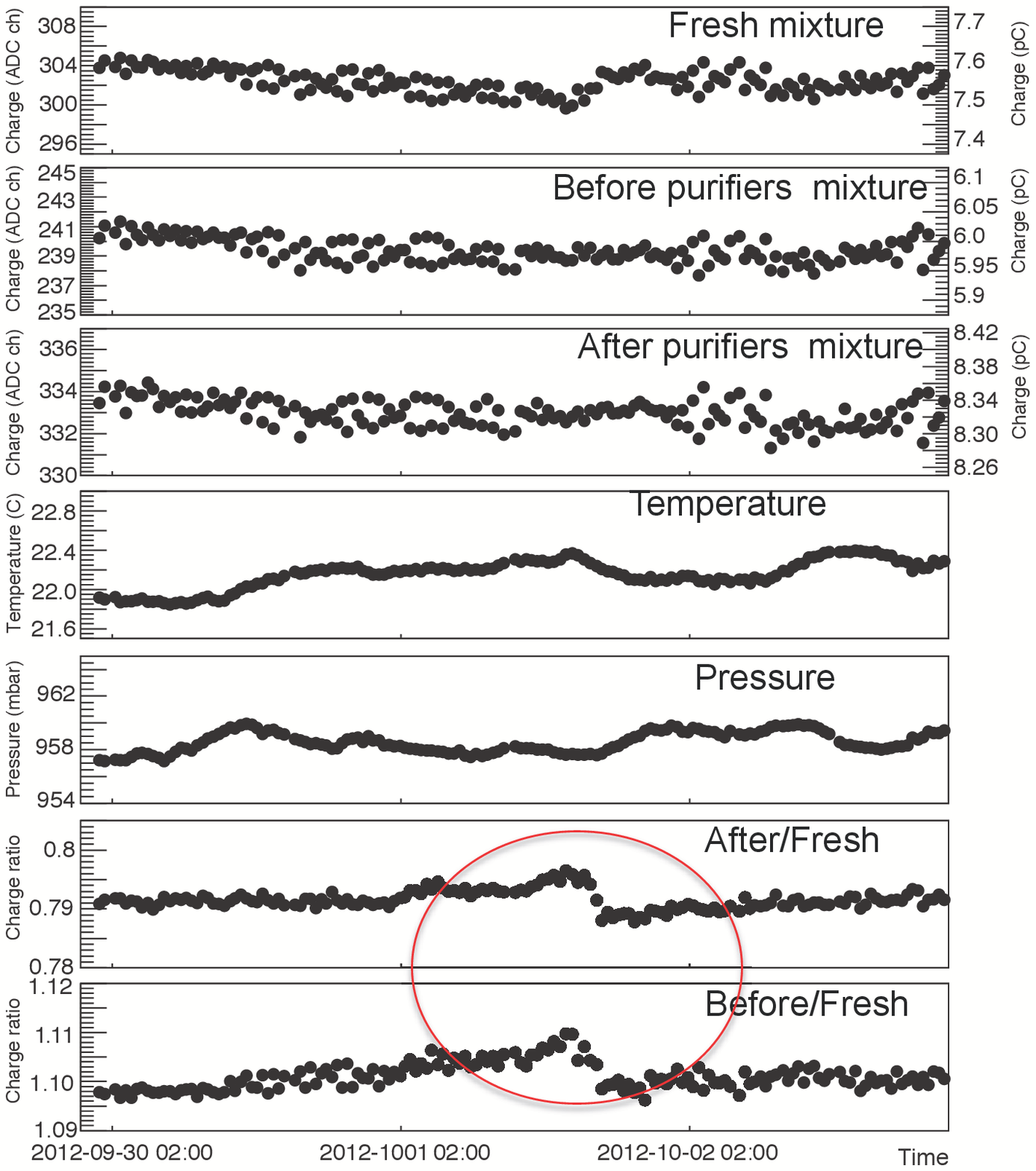}
	\caption{Analysis of the transient interference pattern and correlation with charges and
	environmental parameters.}
	\label{fig:TRANSIENT}
	\end{figure}

	An unexpected feature was observed thanks to a failure
	 of one of the mass flow controllers (MFC) of the gas mixer 
	 (Fig.~\ref{fig:TRANSIENTONLINE}). The duration of each MFC failure was estimated in
	 approximately 30 minutes. The pattern observed in all charge ratios
	 is interpreted as the interference of the 
	working point changes due to the gas mixture
	 change, convoluted with the transit time of the gas 
	mixture inside the closed-loop
	 recirculation system. The fresh gas mixture reaches the GGM before both the
	 affected gas mixtures from before and after the purifiers in the closed-loop system, thus causing the interference pattern shown in
	 Fig.~\ref{fig:TRANSIENT}. 
	
	\section{Conclusions}\label{sec:CONCLU}
	Preliminary results on the operational experience of the GGM system during the CMS 2011-2012 data
	taking period were given.  Weekly HV scans provide direct measurements of the GGM working point.
	When corrected for common environmental effects, the system shows a stability  
	in working point at the level of 20V over nearly two years. 
	A failure in one of the MFC provided hints to define a tool for the fast monitoring of the gas
	mixture composition. If confirmed, the gas mixture composition might be monitored by the time
	distribution pattern of the fresh-, before-, and after-purifier  
	type gas mixture average charge ratios. A simulation analysis is in progress to
	verify such an interpetation.
\end{document}